\documentclass[a4paper,fleqn,usenatbib,useAMS]{mnras}

\usepackage{graphicx}	
\usepackage{amsmath}	
\usepackage{amssymb}	
\usepackage{multicol}        
\usepackage{bm}		
\usepackage{pdflscape}	
\usepackage{color}
\usepackage{graphicx,dblfloatfix}
\usepackage{subfigure}
\usepackage{lscape}
\usepackage{booktabs}

\newcommand{\bfm}[1]{\mbox{\boldmath{$#1$}}}

\usepackage[T1]{fontenc}
\usepackage{ae,aecompl}

\usepackage{newtxtext,newtxmath}

\title[Dynamics of Didymos after the primary's reshaping]{Constraints on the perturbed mutual motion in Didymos due to impact-induced deformation of its primary after the DART impact}

\author[Hirabayashi et al. (2017)]
{Masatoshi Hirabayashi$^{1}$\thanks{Contact e-mail: \href{mailto:thirabayashi@auburn.edu}{thirabayashi@auburn.edu}}\thanks{Present address: Auburn University, 319 Davis Hall, Auburn, AL 36849, U.S.A.}, 
Stephen R. Schwartz$^{2}$,
Yang Yu$^{3}$,
Alex B. Davis$^{4}$, 
\newauthor
Steven R. Chesley$^{5}$,
Eugene G. Fahnestock$^{5}$, 
Patrick Michel$^{6}$,
Derek C. Richardson$^{7}$,
\newauthor
Shantanu P. Naidu$^{5}$,
Daniel J. Scheeres$^{4}$,
Andrew F. Cheng$^{8}$, 
Andrew S. Rivkin$^{8}$,
\newauthor
Lance A. M. Benner$^{5}$ 
\\ \\
$^{1}$Purdue University, 550 Stadium Mall Drive, West Lafayette, IN 47907, U.S.A., \\
$^{2}$Arizona State University, Tempe, AZ 85287, U.S.A., \\
$^{3}$Beihang University, Beijing 100191, China, \\
$^{4}$The University of Colorado, Boulder, CO 80309, U.S.A., \\
$^{5}$Jet Propulsion Laboratory, California Institute of Technology, Pasadena, CA 91109, U.S.A., \\
$^{6}$Laboratoire Lagrange, Universit{\'e} C{\^o}te d'Azur, Observatoire de la C{\^o}te d'Azur, CNRS, CS 34229, 06304 Nice Cedex 4, France, \\
$^{7}$The University of Maryland, College Park, MD 20742, U.S.A., \\
$^{8}$Applied Physics Laboratory/The Johns Hopkins University, Laurel, MD 20723, U.S.A.}

\date{Last updated 2017 August 6}

\pubyear{2017}

\begin{document}
\label{firstpage}
\pagerange{\pageref{firstpage}--\pageref{lastpage}}
\maketitle

\begin{abstract}
Binary near-Earth asteroid (65803) Didymos is the target of the proposed NASA Double Asteroid Redirection Test (DART), part of the Asteroid Impact \& Deflection Assessment (AIDA) mission concept. In this mission, the DART spacecraft is planned to impact the secondary body of Didymos, perturbing mutual dynamics of the system. The primary body is currently rotating at a spin period close to the spin barrier of asteroids, and materials ejected from the secondary due to the DART impact are likely to reach the primary. These conditions may cause the primary to reshape, due to landslides, or internal deformation, changing the permanent gravity field. Here, we propose that if shape deformation of the primary occurs, the mutual orbit of the system would be perturbed due to a change in the gravity field. We use a numerical simulation technique based on the full two-body problem to investigate the shape effect on the mutual dynamics in Didymos after the DART impact. The results show that under constant volume, shape deformation induces strong perturbation in the mutual motion. We find that the deformation process always causes the orbital period of the system to become shorter. If surface layers with a thickness greater than $\sim 0.4$ m on the poles of the primary move down to the equatorial region due to the DART impact, a change in the orbital period of the system and in the spin period of the primary will be detected by ground-based measurement. 
\end{abstract}





\section{Introduction}
\label{Sec:Introcution}
The proposed NASA Double Asteroid Redirection Test (DART) mission, part of the Asteroid Impact \& Deflection Assessment (AIDA) mission concept, plans to launch spacecraft in March 2021 to target the binary Near-Earth asteroid (65803) Didymos in October 2022 \citep{Cheng2015, Cheng2016, ChengLPSC2017}. In this mission, the DART spacecraft will impact the secondary of Didymos. The primary goal of this mission is to demonstrate a measurable deflection of the orbit of the secondary. If possible, the momentum transfer coefficient, $\beta$, \citep{Holsapple2012} will also be estimated. 

Radar and lightcurve observations have shown the physical properties of Didymos \citep[see details in][]{Michel2016,Naidu2016}. The reported total mass of this system is $5.278 \pm 0.54 \times 10^{11}$ kg, and the bulk density is $2100 \pm 630$ kg/m$^3$. For the primary, the mean diameter is $0.780 \pm 0.078$ km, and the spin period is 2.26 hr. The shape looks like a spherical body with an equatorial ridge, which is a so-called top-shape. For the secondary, the mean diameter is $0.163 \pm 0.018$ km. The distance between the centers of mass of these two objects is $1.18 + 0.04 / - 0.02$ km. The eccentricity of the mutual orbit is less than 0.03, and the orbital period is 11.920 hr. Further details of the physical properties of this system are referred to Tables 4 and 5 in \cite{Michel2016}. In the following discussions, we adopt the nominal values of the estimated properties above for our numerical calculations (Table \ref{Table:PhicalProp}). 

Based on the current spin period, the primary appears to be spinning near its spin limit, which is about 2.2 hr \citep{Warner2009}. Possible failure modes may be landslides \citep{Walsh2008, Scheeres2015landslide} or shape deformation \citep{Hirabayashi2014}. \cite{Hirabayashi2015MNRAS} and \cite{Zhang2017} argued that these failure modes depend on different internal structures. After the DART impact on the secondary, some materials ejected from the impact site eventually reach the primary. \cite{Yu2017} showed that the majority of the primary's surface may be overlaid with the ejecta because the motion of the ejecta is sensitive to the primary's gravity \citep{Dell2016}. \cite{Scheeres2017ACM} stated that shape deformation\footnote{He used `reshaping' to describe the same sense of `shape deformation' in our study.} induces energy transfer from the self-potentials of the bodies into orbital energy. This process causes a change in the orbital elements of the system. Therefore, if a cluster of ejecta landing on the primary has enough kinetic energy, the particles landing on (or hitting with high speed) the surface may induce structural failure, causing the primary to deform permanently. 

In this study, we investigate possible responses of the mutual motion in the system to shape deformation of the primary after the DART impact. We develop a numerical model for simulating the mutual motion in the Didymos system by using a radar shape model of the primary \citep{Naidu2016}. We show that shape deformation of the primary causes a change in the gravity field, perturbing the mutual motion in the system. Note that this work was originally presented at the Lunar Planetary Science Conference in 2017 \citep{Hirabayashi2017LPSC} and is extended to strengthen our discussion in this paper. 

This paper consists of two parts. First, we review the deformation modes of a top-shape asteroid rotating near its spin limit and hypothesize a possible deformation process of the primary after the DART impact (Section \ref{Sec:deform}). Second, we introduce a numerical model based on the full-two body problem \citep{Scheeres2006} and show how the mutual motion is perturbed due to differently deformed shapes of the primary (Section \ref{Sec:Dynamics}).  

\begin{table}
\caption{Physical properties used in our numerical exercises. We do not describe the spin period of the secondary because this quantity does not appear in our analysis. Also, the orbital distance described in the last row is the distance of the centers of mass between the primary and the secondary. }
\label{Table:PhicalProp}
\begin{center}
  \begin{tabular}{ l | l l l}
    \hline
    & & Primary & Secondary \\ \hline
    Mass & kg & $5.12 \times 10^{11}$ & $4.76 \times 10^{9}$ \\
    Bulk density & kg/m$^3$ & $2100$ & 2100 \\ 
    Volume & km$^3$ & $2.44 \times 10^{-1}$ & $2.26 \times 10^{-3}$ \\
    Mean radius & km & $3.87 \times 10^{-1}$ & $8.15 \times 10^{-2}$ \\ 
    Spin period & hr & 2.26 & - \\ 
    Orbital distance & km & - & 1.183 \\ 
    \hline
  \end{tabular}
\end{center}
\end{table}

\section{Possible structural behavior of the primary after the DART impact}
\label{Sec:deform}
\subsection{Sensitivity to shape deformation of the primary due to rotation}
\label{Sec:structure}
The primary structurally fails due to fast rotation when its material strength is not strong enough to hold the shape of the original body. The current spin period of the primary (2.26 hr) is close to the spin barrier of asteroids, $\sim 2.2$ hr, \citep{Warner2009}. It is top-shaped with flat poles and an equatorial ridge \citep{Naidu2016}. \cite{Zhang2017} used a Soft Sphere Discrete Model (SSDEM) to conduct comprehensive analyses for the failure mode of the primary of Didymos under the assumption that it is cohesionless. Taking into account possible structural configurations of the primary, they investigated the failure condition and deformation mode of the primary. They concluded that the current shape might be close to its failure condition, depending on the bulk density and size within observation error. If the bulk density were to be lower than the nominal value, even higher values of either one or both the angle of friction or cohesion would be required to maintain the current shape \citep{Zhang2017}. 

\begin{figure*}
  \centering
  \includegraphics[width=5in]{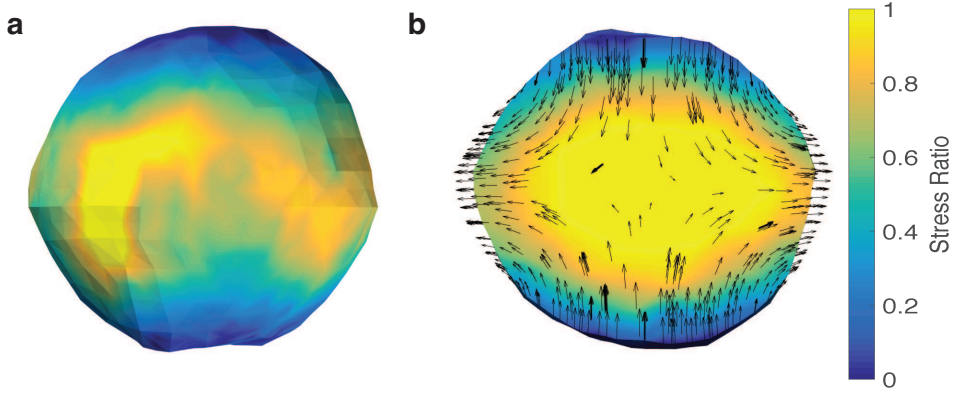}
  \caption{Failure mode of the primary at the present spin period (2.26 hr). The spin axis is along the vertical direction. The contour color shows the stress ratio. When the stress ratio is unity, the region should fail. a. The stress ratio distribution on the surface. b. The stress ratio distribution across the cross section. The arrows describe the total deformation vectors.} 
  \label{Fig:Didymos_failure}  
\end{figure*}

Here we use a plastic finite element model (FEM) technique developed by \cite{Hirabayashi2016Nature} to show an example of the failure mode of the primary due to fast rotation for the case when the material distribution in the primary is uniform. The detailed descriptions for mesh development, boundary conditions, and loading settings are described in \cite{Hirabayashi2016Nature}. We use the radar shape model developed by \cite{Naidu2016}. Figure \ref{Fig:Didymos_failure} shows the failure mode of the primary. For the physical properties, see Table \ref{Table:PhicalProp}. To derive this stress solution, we also use shear strength parameters for geological materials, the friction angle and cohesive strength. The friction angle is an angle describing friction, while the cohesive strength is shear strength at zero pressure \citep{Lambe1969soil}. The friction angle and cohesive strength are fixed at 35$^\circ$ and 25 Pa, respectively, implying that the actual values of these quantities should be higher because its shape is not failing at present.  

The spin axis aligns vertically, and the centrifugal forces always act on the body in the horizontal direction. The contour describes the ratio of the actual stress state to the yield stress state, which is called the stress ratio. If this ratio becomes unity, the elements structurally fail. Figure \ref{Fig:Didymos_failure}a indicates the stress ratio on the surface while Figure \ref{Fig:Didymos_failure}b displays that of a vertical cross section through the pole. It is found that the unity stress ratio spreads over the internal region, meaning that internal failure induces the primary's deformation.

For this case, the failure mode is composed of horizontally outward deformation on the equatorial plane and inward deformation in the vertical direction (see the arrows in Figure \ref{Fig:Didymos_failure}b). These two modes result from an interior that is more sensitive to structural failure than the surface region, making the primary more oblate. The failure mode of this body is comparable to that of Bennu \citep{Scheeres2016Bennu}. The settings of the friction angle and cohesive strength show that the shear resistance of the primary is comparable to that of 1950 DA \citep{Rozitis2014, Hirabayashi2014}, implying that the primary might be close to its failure condition. Also, while in this study we consider the primary to have cohesion, the obtained failure mode is consistent with that derived in \cite{Zhang2017}. 

Other deformation modes may be possible. A landslide has been proposed to be a deformation mechanism \citep{Walsh2008, Walsh2012Spin, Scheeres2015landslide}. For this case, the primary's structure should have a strong interior. As the spin period becomes shorter, the internal structure can still remain intact while the surface layer structurally fails due to stronger centrifugal forces \citep{Hirabayashi2015Internal, Hirabayashi2015MNRAS, Zhang2017}. \cite{Statler2014} also proposed that the Coriolis force may cause deflection of a landslide flow towards the longitude direction, enhancing asymmetric features of the shape. On the other hand, if the interior is structurally weak, the failure mode is characterized by substantial deformation in which the surface layer squashes the internal region, possibly causing a bilobate structure \citep{Sanchez2015DPS}. 

\subsection{Possible deformation path after the DART impact}
\label{Sec:ShapeDeform}
If there is no disturbance to the Didymos system, the shape of the primary should not change. However, since the primary may be sensitive to structural failure, perturbations due to added kinetic energy may trigger shape deformation. The DART impact generates materials ejected from the secondary's surface, some of which arrive at the primary \citep{Yu2017}. If the ejected particles have enough kinetic energy, it is possible that impacts of these particles on the primary induce shape deformation of the primary at any possible scales. Seismic shaking may be a possible factor that could change the surface topography \citep{Richardson2004, Richardson2005} although wave attenuation might be critical in highly porous media. \cite{Murdoch2017} propose that low-energy impacts fluidize more granular materials in a low-gravity environment than those in a terrestrial environment. 

Answering whether or not the DART impact causes the primary's shape deformation requires sophisticated investigation tools for analyzing the impact processes of multiple particles in a low-gravity environment. We leave this investigation as our future work. Here, we consider a possible deformation path of the primary, assuming that the primary deforms due to the process described in Section \ref{Sec:structure}. If granular materials in the primary are fluidized by impacts, strong centrifugal forces may contribute to the deformation process. The primary deforms until the configuration settles into a new equilibrium under constant angular momentum \citep{Holsapple2010}. 

\begin{figure}
  \centering
  \includegraphics[width=3in]{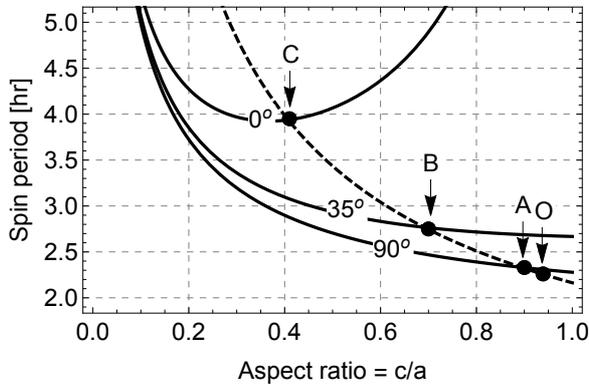}
  \caption{A possible deformation path of the primary. The $x$ axis is the aspect ratio while the $y$ axis is the spin period. The deformation path under constant angular momentum (the dashed line) is plotted on the the equilibrium shape map of Holsapple (the solid lines). To create the equilibrium shape map, we assume that the shape is perfectly oblate, and the structure is cohesionless. For the solid lines from top to bottom, the friction angles are $0^\circ$, $35^\circ$, and $90^\circ$. Location O is the current aspect ratio at the present spin period. Locations A, B, and C describe the aspect ratios at the intersections between the deformation path and the equilibrium shape curves.} 
  \label{Fig:Didymos_deform}  
\end{figure}

Figure \ref{Fig:Didymos_deform} shows a derived deformation path of the primary (the dashed line) and equilibrium curves with different friction angles for an oblate cohesionless body, which are given by \cite{Holsapple2001} (the solid curves). The $x$ axis is the aspect ratio, i.e., the ratio of the semi-minor axis to the semi-major axis of the primary, and the $y$ axis is the spin period. To compute the equilibrium shape curves, we use the defined physical properties (Table \ref{Table:PhicalProp}). The friction angles of the equilibrium shape curves are 0$^\circ$ for the top curve, 35$^\circ$ for the intermediate curve, and 90$^\circ$ for the bottom curve \citep[see the details in][]{Holsapple2001}. 

\begin{figure*}
  \centering
  \includegraphics[width=\textwidth]{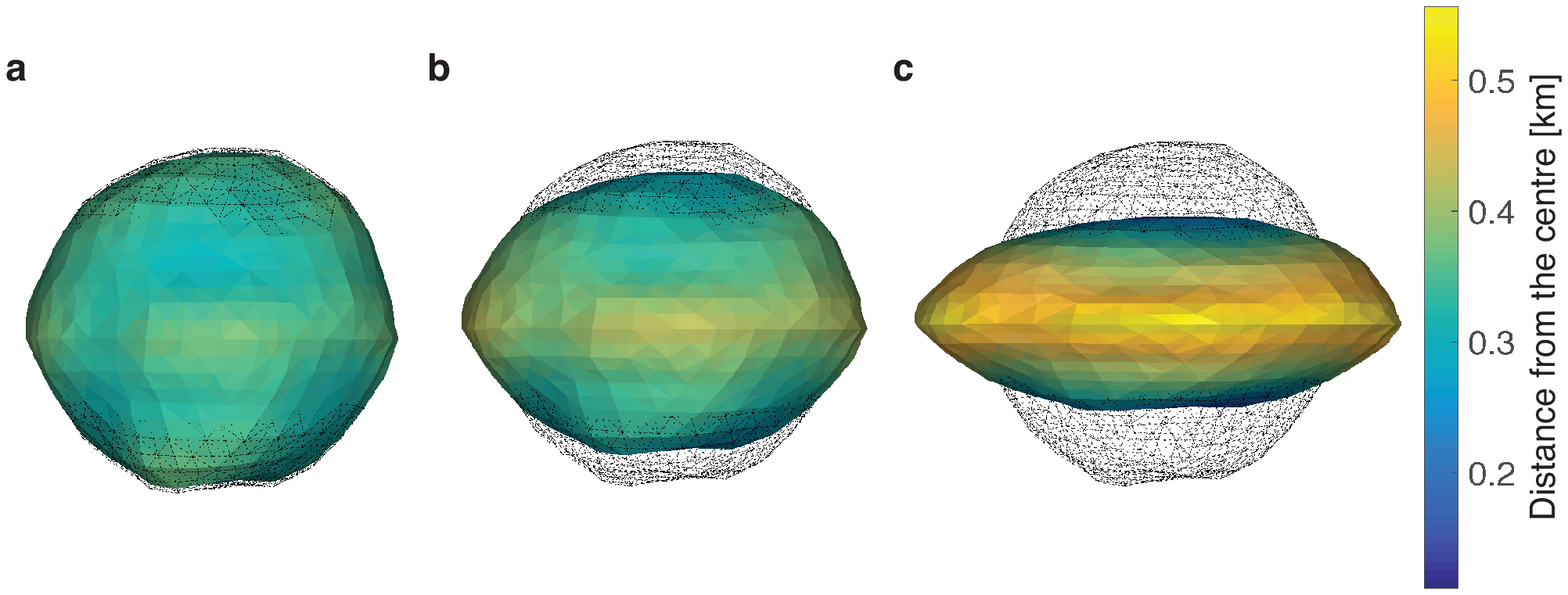}
  \caption{Deformed shapes of the primary. The contour displays the distance of the surface element from the center of mass, and the dots show the original shape. a. Deformed shape at location A in Figure \ref{Fig:Didymos_failure}. b. Deformed shape at location B. c. Deformed shape for location C.} 
  \label{Fig:Didymos_shape}  
\end{figure*}

We derive the deformation path by considering how oblate the original shape becomes under constant volume and angular momentum. Depending on the magnitude of the deformation process, the primary's shape configuration moves on the deformation path towards a lower friction angle. The current aspect ratio is 0.939, which is located at O in Figure \ref{Fig:Didymos_deform}, and is computed based on radar-derived dimensions with a 3$\sigma$ uncertainty of 10 $\%$. The aspect ratios on the deformation path at friction angles of 90$^\circ$, 35$^\circ$, and 0$^\circ$ are 0.9, 0.7, and 0.4, respectively. These locations are denoted as A, B, and C, respectively. Figure \ref{Fig:Didymos_shape} displays the deformed shapes. Note that the case of a friction angle of 0$^\circ$ may be extreme. Also, for real soils, dilatancy increases the volume due to shear, given the initial condition of porosity \citep{Holsapple2010}. Thus, our volume-constant assumption would give conservative results of the orbital perturbation of Didymos; in other words, because our predicted oblateness would be less than the actual one, our derived orbital perturbation might be underestimated. 

These figures show that if the primary is cohesionless, the current configuration has to be supported by an extremely high friction angle; however, a smaller volume within observation error may allow the primary to keep the original shape without cohesion \citep{Zhang2017}. The deformed shapes on the path will be used to evaluate the perturbation of the mutual orbit due to shape deformation of the primary in the following sections. Since it is unknown how the aspect ratio evolves on the deformation path, we choose these four aspect ratio, i.e., 0.939, 0.9, 0.7, and 0.4, as sample cases. 



\section{Dynamical behavior of the system}
\label{Sec:Dynamics}
\subsection{Modeling of dynamical motion}
\begin{figure}
  \centering
  \includegraphics[width=3in]{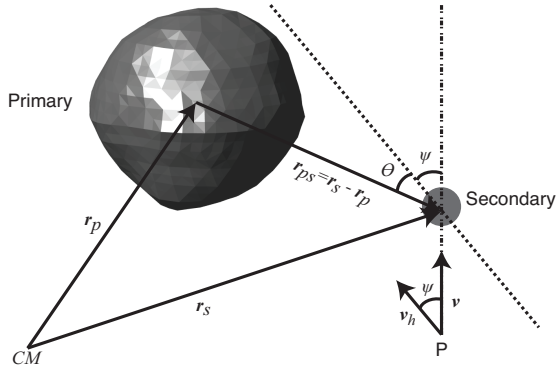}
  \caption{Description of the Didymos system. The primary is modeled using the radar shape model while the secondary is assumed to be a spherical body. The dot-dashed line shows the path of the DART spacecraft while the dashed line describes that projected onto the secondary's orbital plane. $CM$ is the center of mass of the system. $P$ indicates the direction from which the DART spacecraft approaches the secondary. $\theta$ is the phase angle between the approaching path projected on the secondary's orbital plane and the line along the two objects, and $\psi$ is the out-of-plane angle between the velocity vector of the DART spacecraft and the orbital plane. $\bfm v$ is the incident velocity vector while $\bfm v_h$ is the velocity vector projected onto the secondary's orbital plane. }
  \label{Fig:Schematics}  
\end{figure}

We model the mutual interaction between the primary and the secondary using the full two-body problem technique developed by \cite{Scheeres2006}. We use the radar shape model \citep{Naidu2016} for the primary and assume the secondary to be spherical. Note that while full interactions between irregular bodies have been modeled \citep{Werner2005,Fahnestock2006,Hirabayashi2013,Naidu2015,Hou2016,Davis2017}, ours may be a reasonable assumption as radar observations have given few constraints on the secondary's shape.\footnote{Taking into account tidal forces, \cite{Michel2016} assumed the shape of the secondary to be an ellipsoidal shape.} Figure \ref{Fig:Schematics} describes the system. The origin, denoted as $CM$, corresponds to the center of mass of the system. The position vectors of the primary and the secondary are denoted as $\bfm r_{p}$ and $\bfm r_s$, respectively. The relative position of the secondary with respect to the primary is given as $\bfm r_{ps} = \bfm r_{s} - \bfm r_{p}$. The mass and the volume are defined as $M_i$ and $V_i$, respectively, where $i = (p, \: s)$. The dot-dashed line indicates the path of the DART spacecraft, which is supposed to approach the secondary from direction $P$. The trajectory of the DART spacecraft projected onto the secondary's orbital plane is given as the dashed line. We use $\psi$ and $\theta$ to define the impact location. $\psi$ is the angle between the velocity vector of the DART spacecraft and the secondary's orbital plane while $\theta$ is the phase angle indicating the secondary's location on its orbital plane at time of impact.

We describe the mutual gravity force acting on the secondary from the primary as \citep{Scheeres2006}
\begin{eqnarray}
\bfm f_{ps} = - G M_s \int_{V_p} \frac{\bfm r_{ps} - \delta \bfm r_p}{\| \bfm r_{ps} - \delta \bfm r_p \|^3} dM_p = - M_s \frac{\partial U}{\partial \bfm r_{ps}}, 
\end{eqnarray}
where $G$ is the gravitational constant, and $\delta {\bfm r}_p$ is the position vector of an element in the primary relative to the primary's center of mass. Bold letters indicate vectors. We obtain the equation of motion as 
\begin{eqnarray}
&& \ddot {\bfm r}_{ps} + 2 \bfm \Omega_p \times \dot {\bfm r}_{ps} + \dot {\bfm \Omega}_p \times \bfm r_{ps} + \bfm \Omega_p \times (\bfm \Omega_p \times {\bfm r}_{ps}) \nonumber \\
&& = - \left( 1 + \frac{M_s}{M_p} \right) \frac{\partial U}{\partial \bfm r_{ps}},
\end{eqnarray}
where ${\bfm \Omega}_p$ is the spin vector of the primary, and the dots on letters define time-derivatives in the frame rotating with the primary. We also describe the attitude motion of the primary. The torque acting on the primary is given as
\begin{eqnarray}
\bfm \tau_{p} &=& M_s \bfm r_{ps} \times \frac{\partial U}{\partial \bfm r_{ps}}.
\end{eqnarray}
Then, the attitude motion of the primary is given as 
\begin{eqnarray}
\bfm I_p \dot {\bfm \Omega}_p + \bfm \Omega_p \times \bfm I_p \bfm \Omega_p &=& {\bfm \tau}_p. 
\end{eqnarray}
Note that the secondary is spherical, its attitude motion can be decoupled from the mutual dynamics considered here. 

To describe the rotation of the primary, we use Euler parameters \citep{Schaub2003}
\begin{eqnarray}
\beta_0 &=& \cos \frac{\phi}{2}, \\
\beta_1 &=& e_1 \sin \frac{\phi}{2}, \\
\beta_2 &=& e_2 \sin \frac{\phi}{2}, \\
\beta_3 &=& e_3 \sin \frac{\phi}{2},
\end{eqnarray}
where $\bfm e = (e_1, e_2, e_3)^T$ is the principal rotation vector and $\phi$ is the principal angle. These parameters should satisfy
\begin{eqnarray}
\beta_0^2 + \beta_1^2 + \beta_2^2 + \beta_3^3 = 1. 
\end{eqnarray}
The rates of Euler parameters are given as
\begin{eqnarray}
\begin{bmatrix}
\dot \beta_0 \\ \dot \beta_1 \\ \dot \beta_2 \\ \dot \beta_3
\end{bmatrix}
= 
\frac{1}{2}
\begin{bmatrix}
\beta_0 & - \beta_1 & - \beta_2 & - \beta_3 \\
\beta_1 &   \beta_0 & - \beta_3 &   \beta_2 \\
\beta_2 &   \beta_3 &   \beta_0 & - \beta_1 \\
\beta_3 & - \beta_2 &   \beta_1 &   \beta_0 \\
\end{bmatrix}
\begin{bmatrix}
0 \\ \Omega_1 \\ \Omega_2 \\ \Omega_2
\end{bmatrix}.
\end{eqnarray}
Here, $\bfm \Omega_p = (\Omega_1, \Omega_2, \Omega_3)^T$ is described in the rotating frame. Then, the transformation matrix from the inertial frame to the rotating frame, $\bfm A$, is given as
\begin{small}
\begin{eqnarray}
\bfm A = 
\begin{bmatrix}
\beta_0^2 + \beta_1^2 - \beta_2^2 - \beta_3^2 & 2 (\beta_1 \beta_2 + \beta_0 \beta_3) & 2 (\beta_1 \beta_3 - \beta_0 \beta_2) \\
2 (\beta_1 \beta_2 - \beta_0 \beta_0) & \beta_0^2 - \beta_1^2 + \beta_2^2 - \beta_3^2 & 2 (\beta_2 \beta_3 + \beta_0 \beta_1) \\
2 (\beta_1 \beta_3 + \beta_0 \beta_2) & 2 (\beta_2 \beta_3 - \beta_0 \beta_1) & \beta_0^2 - \beta_1^2 -\beta_2^2 + \beta_3^2 
\end{bmatrix}.
\end{eqnarray}
\end{small}

For this system, there are four integrals \citep{Scheeres2006}. The first integral is the total energy of the system, which is given as
\begin{eqnarray}
E &=& \frac{1}{2} \frac{M_p M_s}{M_p + M_s} (\dot {\bfm r}_{ps} + {\bfm \Omega}_p \times {\bfm r}_{ps}) \cdot (\dot {\bfm r}_{ps} + {\bfm \Omega}_p \times {\bfm r}_{ps}) \nonumber \\
&+& \frac{1}{2} {\bfm \Omega}_{p} \cdot {\bfm I}_p {\bfm \Omega}_p + M_s U. 
\end{eqnarray}
The other integrals are defined by the total angular momentum, for which the vector expression is 
\begin{eqnarray}
{\bfm H} = \bfm A \left[ {\bfm I}_p {\bfm \Omega}_p +  \frac{M_p M_s}{M_p + M_s} \left \{ {\bfm r}_{ps} \times \dot {\bfm r}_{ps} + {\bfm r}_{ps} \times ({\bfm \Omega}_p \times {\bfm r}_{ps}) \right \} \right]. 
\end{eqnarray}

\subsection{Results}
In this section, we explore how the deformed primary changes the mutual motion of Didymos after the DART impact. The integrator used is an eighth-order Runge-Kutta scheme \citep{Montenbruck2000} with a fixed step size of 432 sec. We assume that the initial mutual orbit between the primary and the secondary has zero inclination and zero eccentricity. We also consider that the DART impact is at $27.5^\circ$ out of the secondary's orbital plane \citep{Cheng2016}, i.e., $\psi = 27.5^\circ$. Again, the physical property used are given in Table \ref{Table:PhicalProp}. 

\begin{figure}
  \centering
  \includegraphics[width=3in]{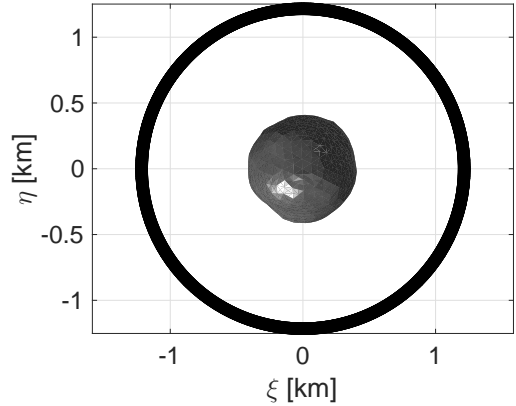}
  \caption{Mutual motion of the Didymos system that is given in the rotating frame fixed to the primary. The origin is identical to the center of mass of the primary. The coordinate frame is fixed to the primary. The body drawn at the center is the primary. The thick circle is the secondary's orbit, which is perturbed by the primary. } 
  \label{Fig:Didymos_nominal}  
\end{figure}

We first compute the nominal case in which the primary does not deform. Figure \ref{Fig:Didymos_nominal} shows the orbit of the secondary relative to that of the primary in the coordinate frame fixed to the primary. The $\xi$, $\eta$, and $\zeta$ axes are defined along the the primary's minimum, intermediate, and maximum moment of inertia axes, respectively. This figure plots the motion in the $\xi - \eta$ plane. The thick, circular orbit indicates that the shape of the primary perturbs the mutual orbit between the primary and the secondary. Next, we investigate the perturbation of the mutual orbit caused by the DART impact. We consider that at $t = 0$ the impact process instantaneously changes the velocity of the secondary. For simplicity, we consider that the momentum transfer coefficient is at one; in order words, we do not account for added momentum transfer from ejecta in this analysis. Given the linear momentum conservation, we compute the change in the velocity of the secondary as 
\begin{eqnarray}
\Delta {\bfm v}_s = \frac{m \bfm v}{M_2},
\end{eqnarray} 
where $m$ is the mass of the DART impactor, and $\bfm v$ is the incident velocity vector (Figure \ref{Fig:Schematics}). To model the DART impact, we fix $m$ and $\| \bfm v\|$ at 500 kg and 6 km/s, respectively.\footnote{Theses values are current as of February 17, 2017.} We obtain the change in the speed of the secondary, $\| \Delta {\bfm v}_s \|$, as $6.3 \times 10^{-4}$ m/sec. Because of the DART impact angle, the velocity change of the secondary on the orbital plane is $6.3 \times 10^{-4} \cos (27.5^\circ) = 5.6 \times 10^{-4}$ m/sec.  

\begin{figure*}
  \centering
  \includegraphics[width=\textwidth]{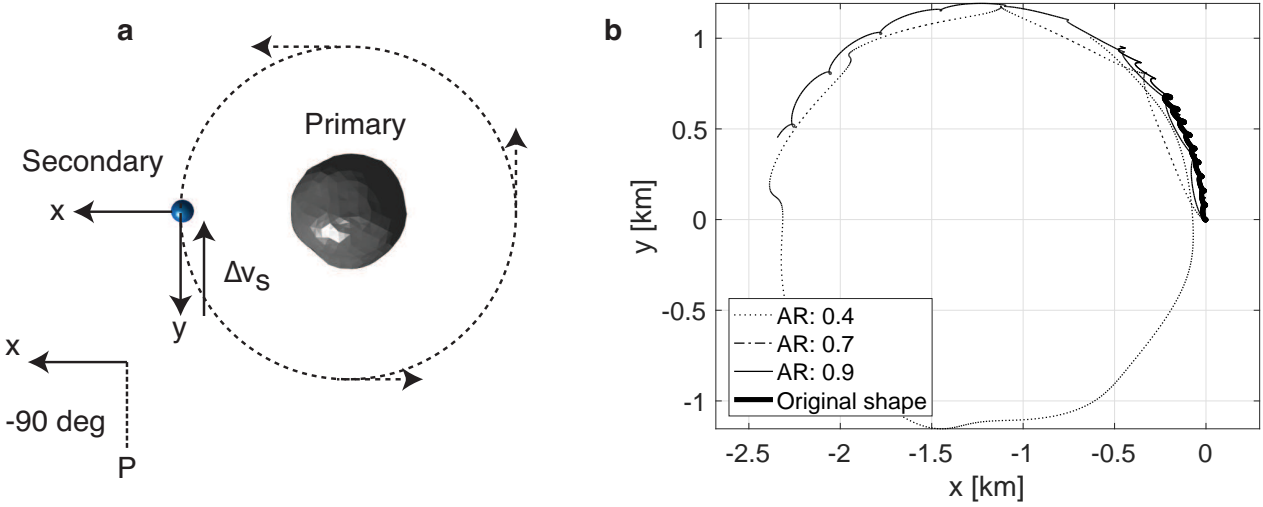}
  \caption{a. Currently planned DART impact location. The phase angle, $\theta$, is 90$^\circ$. The directions of the velocities of the secondary and the DART impactor are opposite. b. The secondary's orbital perturbation from its nominal orbit after the DART impact. The bold line shows the case when the primary does not deform at all. The narrow dotted, dot-dashed, and solid lines describe the cases at aspect ratios of 0.4, 0.7, and 0.9, respectively. $AR$ in the legend stands for `Aspect Ratio.' } 
  \label{Fig:Didymos_case01}  
\end{figure*}

\begin{figure*}
  \centering
  \includegraphics[width=\textwidth]{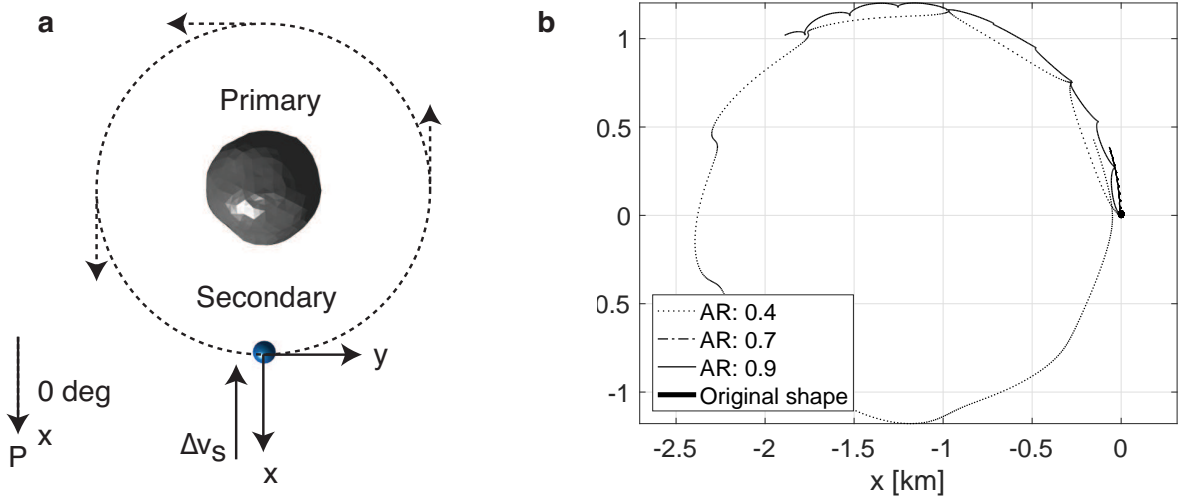}
  \caption{a. Hypothesized impact location. The directions of the velocities of the secondary and the DART impactor are perpendicular to each other. b. The secondary's orbital perturbation from its nominal orbit after the DART impact. The styles of the lines are the same as given in Figure \ref{Fig:Didymos_case01}.} 
  \label{Fig:Didymos_case02}  
\end{figure*}

In this study, we consider the currently planned impact location and an additional test location to show how the initial location affects the orbital evolutions (see panel a in Figures \ref{Fig:Didymos_case01} and \ref{Fig:Didymos_case02}). The currently planned impact location is at $\theta = - 90^\circ$. Note that the exact location may be slightly different from our defined location. The test location is fixed at $\theta = 0^\circ$. This location is less likely to be selected as the impact site because it is difficult to observe the effect of momentum transfer. However, we consider this case to demonstrate that the DART impact at this site may trigger orbital perturbation not by addition of momentum by the DART impact but by shape deformation. In addition, while materials on the secondary would be ejected in the direction opposite to the location of the primary, low-velocity ejecta might be trapped by the primary's gravity, and some of them would still reach the primary. With these initial impact locations, we investigate the effect of the deformed shapes on the perturbation of the mutual orbit, considering the four aspect ratios defined in Section \ref{Sec:ShapeDeform}. 

We calculate the orbital perturbation within 10 orbital periods, equivalent to 4.97 Earth days, for these cases. Panel b in Figures \ref{Fig:Didymos_case01} and \ref{Fig:Didymos_case02} show the secondary's orbit relative to its nominal orbit (see Figure \ref{Fig:Didymos_nominal}). The $x$ axis defines the orbital perturbation from the nominal case in the radial direction, and the $y$ axis gives the deviation in the tangential direction. In other words, these two axes rotate with the nominal location of the secondary. We omit the descriptions of the motion in the out-of-plane direction. Because of this coordinate frame setting, the maximum distance between the nominal location and the perturbed location in the $x$ axis should be two times the orbital radius, i.e., $\sim 2.36$ km, at $y \approx 0$ km, while that in the $y$ axis should be identical to the orbital radius, i.e., $\sim 1.18$ km, at $x \approx -1.18$ km. These features are seen in these plots. Each orbit of the secondary is differently affected by the primary's shape deformation. The bold line shows the secondary's orbit influenced by the DART impact without the primary's deformation. For this case, the aspect ratio of the primary is 0.939. The narrow solid, dot-dashed, and dotted lines describe the orbital motion of the secondary after the DART impact with aspect ratios of 0.9, 0.7, and 0.4, respectively. The origin of the frame is identical to the location of the secondary in the nominal case at a given time. 

The results show that the deformed primary changes the gravity field in the system, affecting the mutual interaction between the primary and the secondary. We first discuss the orbital perturbation after the DART impact at the currently planned location, i.e., $\theta = -90^\circ$ (Figure \ref{Fig:Didymos_case01}). If the primary does not deform at all (the bold black line), the orbital energy of the system decreases due to the kinetic energy of the DART impact, and the distance between the primary and the secondary becomes shorter. Because of this process, the orbital period is 357.6 sec (= 5.96 min) shorter than that for the nominal case (Table \ref{Table:OrbitalPeriod}).\footnote{Our result is 87.6 sec shorter than the value derived by \cite{Cheng2016}. This slight difference comes from the use of the updated spacecraft configurations and the radar shape model.} If the primary deforms, the orbit of the secondary is perturbed by the change in the gravity field (the narrow lines). Since the deformation process always makes the primary's aspect ratio smaller, the gravity force in the radial direction increases on the equatorial plane, pulling the secondary inward. Therefore, similar to the no-deformation case, the orbital period becomes shorter. Depending on an aspect ratio after the deformation process, a change in the orbital period may become significant (Table \ref{Table:OrbitalPeriod}). For the cases of $\theta = 0^\circ$, we find that the orbital perturbation due to the primary's deformation is consistent with the case of $\theta = -90^\circ$ (Figures \ref{Fig:Didymos_case02}). In conclusion, the DART impactor makes the orbital period shorter; likewise, the deformation process of the primary also shortens the orbital period. 

So far, we studied the orbital perturbation, given fixed aspect ratios of the primary's shape. However, since the magnitude of the deformation process is unknown, it is difficult to determine the level of the orbital perturbation. Therefore, we also consider how large the deformation should be to affect ground-based measurement. The mission requirement for measurement accuracy of a change in the orbital period is 7 sec. We determine the aspect ratio of the primary such that the orbital period is 7 sec shorter than that for the case of no deformation. We obtain that the aspect ratio at this condition is 0.938, which could happen if the surface layers with a thickness thicker than $\sim 0.4$ m at the poles move down to the equatorial region. 

While a change in the orbital period should be measured accurately, our results imply that if shape deformation occurs at such a small scale or larger, it is likely to influence momentum transfer estimation planned on the DART mission. Thus, it is vitally important to separate the effect of shape deformation from that of the DART impact. One way of observing this effect might be to observe a change in the spin period of the primary. As shown in Figure \ref{Fig:Didymos_shape}, under constant angular momentum, the spin period may change due to the deformation process. We write the spin period change of the primary as
\begin{eqnarray}
\Delta T = \left(\frac{I_{pz}}{I_{pz0}} - 1 \right) T_0, \label{Eq:DeltaT}
\end{eqnarray}
where $I_{pz0}$ and $I_{pz}$ are the maximum moment of inertia components of the primary before and after the DART impact, respectively, and $T_0$ is the original spin period of the primary, i.e., 2.26 hr. Since $I_{pz} > I_{pz0}$, $\Delta T > 0$; that is, the new spin period is always slower than the original spin. Table \ref{Table:SpinPeriod} shows how $\Delta T$ depends on the final shape. If we observe the spin period change, it is possible to decouple the DART impact effect and the shape deformation effect. Importantly, even if the deformed aspect ratio is 0.938, the spin period change is 5.781 sec (=0.0016 hr), which is still detectable.\footnote{A currently reported observation error of the primary's spin period is 0.0001 hr \citep{Michel2016}.} However, since mutual dynamics of the system is likely to provide critical effects on the primary's spin condition, it is necessary to develop sophisticated analysis tools and observation techniques. 

\begin{table}
\caption{Changes in the orbital period after the DART impact. The units are seconds. The negative values describe that the orbital period becomes shorter than that in the nominal case. For the definition of $\theta$, see Figure \ref{Fig:Schematics}.}
\label{Table:OrbitalPeriod}
\begin{center}
  \begin{tabular}{ l | l l l l}
    \hline
     {} & & \multicolumn{2}{c}{Aspect ratio} & \\
     {} & 0.939 & 0.9 & 0.7 & 0.4 \\ 
    \hline 
    $\theta = - 90^\circ$ & -357.6 & -595.8 &  -1872 & -5004 \\ 
    $\theta =   0^\circ$ & 0 & -238.2 & -1488 & -4530 \\ 
    \hline
  \end{tabular}
\end{center}
\end{table}

\begin{table}
\caption{Changes in the spin period of the primary after the deformation process. $\Delta T$ is defined in Equation (\ref{Eq:DeltaT}). The units are seconds. Note that the original aspect ratio is 0.939. }
\label{Table:SpinPeriod}
\begin{center}
  \begin{tabular}{ l | l l l l}
    \hline
     {} & & \multicolumn{2}{c}{Aspect ratio} & \\
     {} & 0.938 & 0.9 & 0.7 & 0.4 \\ 
    \hline 
    $\Delta T$ & 5.781 & 233.4 & 1760 & 6235 \\ 
    \hline
  \end{tabular}
\end{center}
\end{table}

\section{Discussion \& conclusion}
We investigated how the mutual orbit in binary near-Earth asteroid Didymos would change due to shape deformation of the primary. The primary is currently rotating with a spin period of 2.26 hr, which may be close to its critical spin condition. Since some materials ejected from the secondary by the DART impact reach the primary \citep{Yu2017}, they may affect the sensitivity of the primary to structural failure. Assuming that such a process changes the shape of the primary, we conducted numerical simulations to compute dynamical interaction in the Didymos system after the DART impact. Specifically, we analyzed how mutual motion between the primary and the secondary would evolve due to the primary's deformation. We showed strong perturbation in the system due to the gravity field of the deformed primary under constant volume. As the aspect ratio of the primary decreases due to deformation, the gravity force in the radial direction became larger, making the orbital period shorter. 

We explain the critical assumptions made in this study. First, the shape of the secondary was assumed to be spherical. At present, ground observations have not detected the shape of the secondary. Thus, in the present study, it is reasonable to assume the secondary to be a sphere. However, if the secondary is non-spherical, the secondary's orbit is coupled with its attitude motion. Early studies showed the coupled motion of binary near-Earth asteroid 1999 KW4 \citep{Scheeres2006KW4, Fahnestock2008, Hou2016}. Specifically, accurate description of the mutual motion may require considerations of up to the fourth order of the inertia integrated tensors \citep{Davis2017}. 

Second, we simplified the deformation mode of the primary in the present study. A recent work demonstrated that even if an asteroid has a symmetric shape, the internal heterogeneity could cause asymmetric deformation \citep{Sanchez2016}. Even if the structure is homogeneous, the Coriolis force may change the direction of a landslide flow towards the longitude direction, causing the shape to become asymmetric \citep{Statler2014}. Also, as mentioned by \cite{Yu2017}, the materials ejected from the secondary after the DART impact may reach the majority of the primary's surface with a range of impact velocities. In such a case, some regions may be unaffected by ejecta while other regions may have local deformation modes, causing asymmetric deformation in the primary. In addition, particles that depart from the primary may reach the secondary. Landslides possibly add additional energy to moving particles \citep{Scheeres2015landslide}. For the Didymos system in which the primary is rotating at a spin period of 2.26 hr, this process may provide them enough energy to arrive at the secondary, which makes the present problem more complex. 

We emphasize that we did not conclude that shape deformation of the primary must happen due to collisions of materials ejected from the secondary by the DART impact. Didymos experiences high-speed impacts from micrometeorites frequently, which supports the hypothesis that the current shape is structurally strong enough to resist such impacts. However, the impact flux should significantly increase after the DART impact \citep{Yu2017}, and it is uncertain if the original shape can remain under such a severe condition. We also note that the effect of the momentum transfer on the secondary, i.e., the case of a momentum transfer coefficient being greater than one, is not considered in this study, and it is necessary to quantify this effect. These are open questions, and further investigation is necessary to quantify a possibility of the primary's deformation. 


\section*{Acknowledgements}
\addcontentsline{toc}{section}{Acknowledgements}
M.H. acknowledges ANSYS 17.1 for FEM computation in this project. 









\bsp	
\label{lastpage}
\end{document}